\documentclass[fleqn,usenatbib]{mnras}

\usepackage{newtxtext,newtxmath}

\usepackage[T1]{fontenc}
\usepackage{ae,aecompl}

\usepackage{graphicx}	
\usepackage{amsmath}	
\usepackage{amssymb}	
\usepackage{color}
\usepackage{cprotect}
\definecolor{linkcolor}{rgb}{0,0,0.25}
\hypersetup{
  colorlinks=true,        
  linkcolor=linkcolor,    
  citecolor=linkcolor,    
  filecolor=linkcolor,    
  urlcolor=linkcolor      
}

\makeatletter
\renewcommand{\@printed}{}
\makeatother

\newcommand{\figurename}{Figure}

\newcommand{\patspeed}{41}
\newcommand{\patspeedranerr}{1.5}
\newcommand{\patspeederr}{3}

\newcommand{\gaia}{\emph{Gaia}}

\newcommand{\Gyr}{\ensuremath{\,\mathrm{Gyr}}}
\newcommand{\kpc}{\ensuremath{\,\mathrm{kpc}}}
\newcommand{\kms}{\ensuremath{\,\mathrm{km\,s}^{-1}}}
\newcommand{\kmskpc}{\ensuremath{\,\mathrm{km\,s}^{-1}\,\mathrm{kpc}^{-1}}}

\newcommand{\Ra}{R_a}

\newcommand{\teff}{\ensuremath{T_{\mathrm{eff}}}}
\newcommand{\logg}{\ensuremath{\log g}}

\title[Dynamics, formation, and evolution of the Milky Way's bar]{Life in the fast lane: a direct view of the dynamics, formation, and evolution of the Milky Way's bar}

\author[Bovy et al.]{
Jo Bovy$^{1,2}$\thanks{E-mail: bovy@astro.utoronto.ca~.}, 
Henry~W.~Leung$^1$, 
Jason~A.~S.~Hunt$^2$, 
J. Ted Mackereth$^3$, 
\newauthor D. A. Garc\'{i}a-Hern\'{a}ndez$^{4,5}$, 
\& Alexandre Roman-Lopes$^6$
\\
$^1$ Department of Astronomy and Astrophysics, University of Toronto, 50 St. George Street, Toronto, ON, M5S 3H4, Canada\\
$^2$ Dunlap Institute for Astronomy and Astrophysics, University of Toronto, 50 St. George Street, Toronto, ON M5S 3H4, Canada\\
$^3$ School of Astronomy and Astrophysics, University of Birmingham, Edgbaston, Birmingham, B15 2TT, UK\\
$^4$ Instituto de Astrof\'{i}sica de Canarias, E-38205 La Laguna, Tenerife, Spain\\
$^5$ Universidad de La Laguna, Departamento de Astrof\'{i}sica, E-38206 La Laguna, Tenerife, Spain\\
$^6$ Departamento de F\'{i}sica, Facultad de Ciencias, Universidad de La Serena, Cisternas 1200, La Serena, Chile
}

\date{}

\pubyear{2019}

\begin{document}
\label{firstpage}
\pagerange{\pageref{firstpage}--\pageref{lastpage}}
\maketitle

\begin{abstract}
Studies of the ages, abundances, and motions of individual stars in the Milky Way provide one of the best ways to study the evolution of disk galaxies over cosmic time. The formation of the Milky Way's barred inner region in particular is a crucial piece of the puzzle of disk galaxy evolution. Using data from APOGEE and \gaia, we present maps of the kinematics, elemental abundances, and age of the Milky Way bulge and disk that show the barred structure of the inner Milky Way in unprecedented detail. The kinematic maps allow a direct, purely kinematic determination of the bar's pattern speed of $\patspeed\pm\patspeederr\kmskpc$ and of its shape and radial profile. We find the bar's age, metallicity, and abundance ratios to be the same as those of the oldest stars in the disk that are formed in its turbulent beginnings, while stars in the bulge outside of the bar are younger and more metal-rich. This implies that the bar likely formed $\approx 8\Gyr$ ago, when the decrease in turbulence in the gas disk allowed a thin disk to form that quickly became bar-unstable. The bar's formation therefore stands as a crucial epoch in the evolution of the Milky Way, a picture that is in line with the evolutionary path that emerges from observations of the gas kinematics in external disk galaxies over the last $\approx10\Gyr$.
\end{abstract}


\section{Introduction}

The Milky Way's bar was discovered in the distribution of near-infrared emission \citep{Blitz91a} and in the kinematics of gas \citep{Binney91a} in the central regions. However, the large and variable amount of interstellar extinction \citep{Nataf13a} has made direct studies of the bar difficult in the 30 years since its discovery. Measurements of stars in fields typically away from the Galactic mid-plane to avoid the effects of interstellar extinction demonstrate that the bar dominates the kinematics in the inner few kpc \citep{Shen10a,Ness13b}, leaving little room for the classical bulge component that is expected to result from the hierarchical galaxy formation paradigm \citep{Kormendy04a}. Bars rotate as solid bodies and the bar's angular pattern speed is its most important property aside from its mass for determining its influence on, e.g., the observed perturbed kinematics of stars in the Galactic disk \citep{Antoja18a,Hunt18a,Fragkoudi19a,2019MNRAS.482.1983F} or on the structure of stellar streams in the halo and this structure's interpretation in terms of dark-matter substructure \citep{Pearson17a,Banik19a}.

The distribution of stellar ages and chemical abundances reveals the formation and evolution of the Milky Way's stellar components. Originally thought to be primarily old \citep{Zoccali03a}, spectroscopic observations of stars in the bulge region have revealed a complex mix of populations spanning a wide range of ages, metallicities, and abundance ratios \citep{Bensby13a,Ness13a}. However, directly separating the bar populations from the inner disk, inner halo, and spheroidal bulge components has been difficult using past observations and the age of the bar and the chemical evolution of its stars remains largely unknown. The chemical evolution of galactic components is traced by their distribution in the $([\mathrm{Fe/H}],[\mathrm{O/Fe}])$ plane, because oxygen is mainly produced through type II supernovae that occur soon after the commencement of star formation, while iron has a large contribution from type Ia supernovae, which lag star formation by typically $\approx1\Gyr$. Determinations of detailed abundances of stars in the bulge show that the bulge follows a chemical evolution track similar to that of the local old, thicker disk component \citep[e.g.,][]{Bensby10a}, indicative of a high star-formation efficiency, but it remains uncertain how consistent the bulge and old disk chemical evolution really is.

In this paper, we combine two powerful new data sets---APOGEE \citep{APOGEE} and \gaia\ \citep{GaiaDR2}---to for the first time directly reveal the bar in spatial maps of the kinematics, chemical abundances, and ages of stars in the Milky Way. We demonstrate that these maps allow a direct, purely kinematic determination of the bar's pattern speed and its variation with radius, the bar's radial profile, and the bar's shape using a novel application of the continuity equation. Furthermore, the fact that we clearly see the bar in the abundances and ages of stars lets us unambiguously separate the bar from the inner bulge and disk, which reveals the bar's chemical evolution history.

\section{Data}

\subsection{APOGEE data}

We use data from the 16th data release (DR16) of the Apache Point Observatory Galactic Evolution Experiment (APOGEE; \citealt{APOGEE}). APOGEE is a high-resolution ($R \approx22,500$), high signal-to-noise ratio (typically $\gtrsim100$), near-infrared ($1.5$ to $1.7\mu\mathrm{m}$), large spectroscopic survey \citep{Nidever15a,2017AJ....154...28B,Zasowski17a,APOGEEDR14,Wilson19a}. DR16 contains 473,038 stars observed using the APO 2.5m telescope \citep{Gunn06a} and the DuPont telescope at Las Campanas Observatory (LCO). The dual-hemisphere, full-sky coverage is crucial for mapping the center of the Milky Way and for producing a consistent data set covering the inner bulge to the outer disk.

\begin{figure}
\begin{center}
\includegraphics[width=0.49\textwidth,clip=]{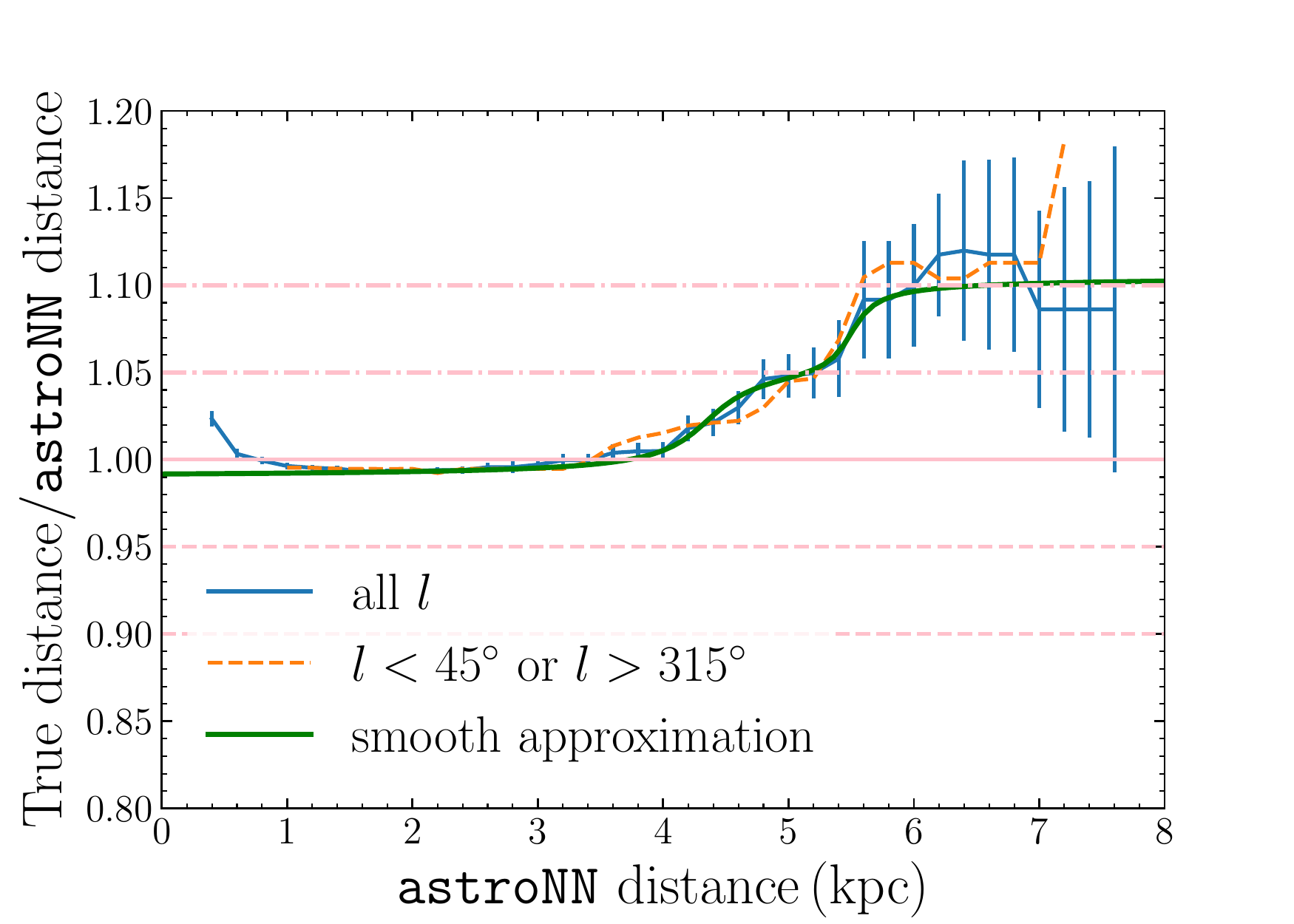}
\end{center}
\caption{Relative distance bias in the raw \texttt{astroNN} distances derived from APOGEE spectra by comparing to \gaia\ parallaxes. We determine the distance systematic by a robust maximum-likelihood stacking of the \gaia\ parallaxes in 200 pc wide bins in raw neural-network distance; the errorbars are derived using bootstrap resampling. The points and errorbars shown are median-smoothed over $1\kpc$ to more clearly display the trend. The blue curve shows the distance bias determined using all stars within 300 pc of the Galactic mid-plane, while the orange curve shows the bias determined using stars towards the direction of the bulge alone, which agrees with the measurement using all stars. The green curve shows a simple  approximation to the bias that we use to correct the distances used in this work.}\label{fig:astronn-dist-systematics}
\end{figure}

We analyze the data using a custom set of analysis pipelines that is based on, but separate from, the normal APOGEE pipeline \citep{Linelist,Zamora15a,ASPCAP}. We start from the one-dimensional, rest-frame corrected, combined-exposure spectra in the \texttt{apStar} files. We perform a (pseudo-)continuum normalization to remove the effect of the overall continuum on stellar abundance, age, and luminosity measurements, by fitting a second-order polynomial to a set of wavelength pixels previously determined to trace the continuum, for each of the three APOGEE detectors separately. To determine abundances, we then run these continuum-normalized spectra through the \texttt{astroNN} trained neural-network \citep{Leung19a} to determine the stellar parameters effective temperature $\teff$ and surface gravity $\logg$ and to determine the elemental abundances of iron Fe, oxygen O, as well as other elements that are not used here. The DR16 spectral reduction imprints an overall difference between the DR16 \texttt{apStar} spectra that we use here and the DR14 spectra that the neural network was trained on. That is, even when the raw data underlying a given \texttt{apStar} spectrum is the same, differences between the DR14 and DR16 spectral reduction cause a small difference in the reduced \texttt{apStar} spectrum. Moreover, this difference is different for DR16 spectra observed at APO and at LCO (DR14 contained only APO observations). To apply the previously trained neural network, we determine the median difference between continuum normalized spectra in DR16 (separately for APO and LCO) using the set of spectra observed both from APO and from LCO to place all spectra onto the same footing. Comparing stellar parameters and Fe and O abundances for stars in the APO/LCO overlap sample shows results consistent between DR14 and DR16 APO (which are the same underlying spectra) to a level of $15\,\mathrm{K}$ in \teff, $0.035\,\mathrm{dex}$ in \logg, $0.015\,\mathrm{dex}$ in $[\mathrm{Fe/H}]$, and $0.018\,\mathrm{dex}$  in $[\mathrm{O/Fe}]$; between spectra of the same star taken at APO and at LCO, differences are at the level of $40\,\mathrm{K}$ in \teff, $0.1\,\mathrm{dex}$ in \logg, $0.04\,\mathrm{dex}$ in $[\mathrm{Fe/H}]$, and $0.05\,\mathrm{dex}$  in $[\mathrm{O/Fe}]$. These differences are smaller or similar to the typical uncertainties in the derived parameters of $60\,\mathrm{K}$ in \teff, $0.2\,\mathrm{dex}$ in \logg, $0.06\,\mathrm{dex}$ in $[\mathrm{Fe/H}]$, and $0.05\,\mathrm{dex}$  in $[\mathrm{O/Fe}]$.

\begin{figure*}
\begin{center}
\includegraphics[width=\textwidth]{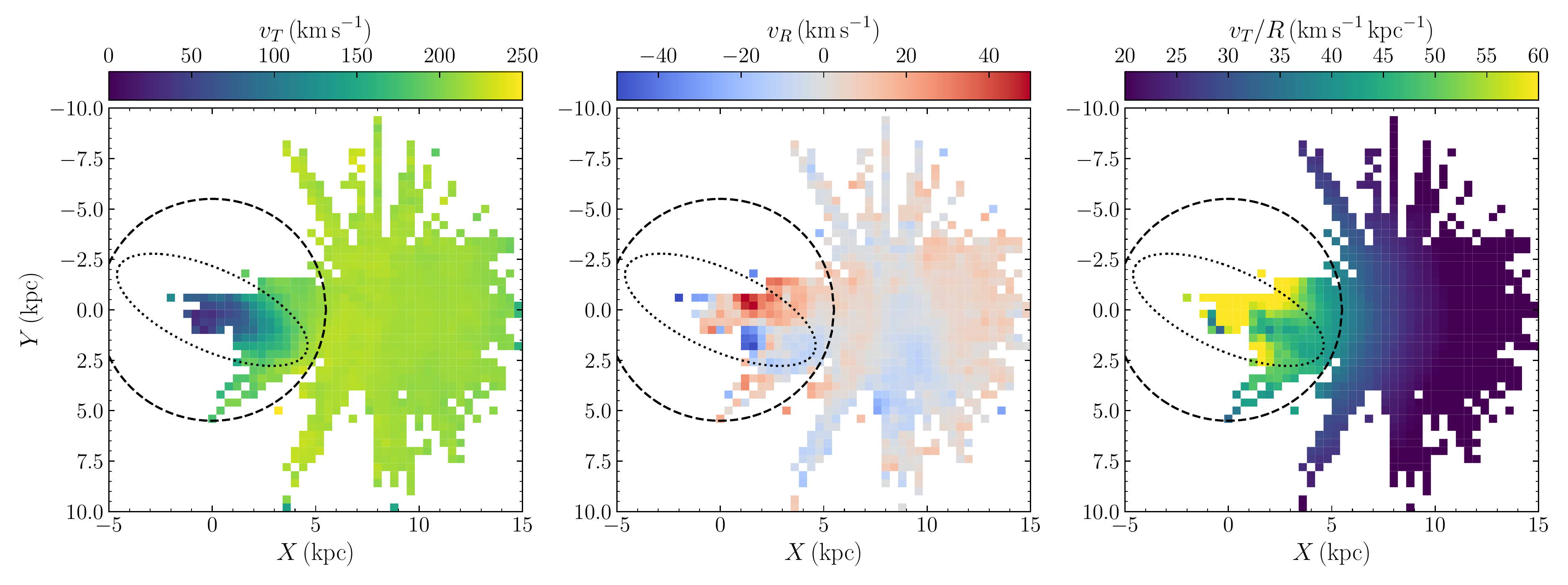}
\end{center}
\caption{Kinematics of stars in the bulge and disk of the Milky Way. The panels display the rotational velocity $v_T$ (\emph{left}), the radial velocity $v_R$ (\emph{center}), and the rotational frequency $v_T/R$ (\emph{right}). A circle with a radius of 5.5 kpc---the bar's corotation radius---and an ellipse oriented at $25^\circ$ clockwise from the Sun--Galactic center line with a semi-major axis of 5 kpc and axis ratio $q=0.4$ are shown for reference. The central bar region is clearly visible in these different kinematic projections with patterns that agree with simulations. The kinematic major axis matches that determined photometrically.}\label{fig:kinematics}
\end{figure*}

We determine distances to the stars in APOGEE DR16 using a previously trained \texttt{astroNN} neural-network \citep{Leung19b} that determines the luminosity of giants from continuum-normalized APOGEE spectra and derives distances from these by combining this with the observed apparent $K_s$ magnitude from 2MASS \citep{2MASS} and extinction values determined using mid-infrared photometry with the RJCE method \citep{RJCE}. Below, we determine that our distances are systematically biased  low by 5\% to 10\% near the mid-plane at distances of 4 kpc and beyond and we correct this bias using its determination below. We convert the 3D heliocentric positions to Galactocentric coordinates by setting the distance to the Galactic center to $8.125\kpc$ \citep{2018A&A...615L..15G} and the Sun's height above the plane to $20.8\,\mathrm{pc}$ \citep{Bennett19a}.

To determine ages, we re-train the neural network that we used to determine ages using DR14 data \citep{Mackereth19a} on the overlap between stars in the APOKASC \citep{APOKASC2} sample with asteroseismic ages and APOGEE DR16, using the DR16-corrected spectra as explained above. As in our previous work, limitations in the training data set make it such that we cannot obtain ages for stars with $[\mathrm{Fe/H}] < -0.5$, but only a small fraction of stars that we consider in the bulge and disk region have such low metallicities and this limitation does not impact our results. As in \citet{Mackereth19a}, the raw ages determined by the neural network at ages $\gtrsim6\Gyr$ are underpredicted, because the reduced sensitivity of carbon and nitrogen spectral features to age at high age causes the predictions from the regularized neural-network used to tend to the mean age of the sample. We measure this bias on the training sample, smoothly interpolate the bias using a Locally Weighted Scatterplot Smoothing (lowess), and apply the lowess bias correction to all predicted ages by the neural network. For our data set, this correction has the effect of shifting all ages towards larger ages, but it has little effect on the comparison of ages of stars in the bar to those of stars outside the bar, because the ages of stars in these different samples are corrected by a similar amount. The typical age precision is $\approx30\,\%$. 

For all kinematic, abundance, and age maps and distributions that we determine, we select stars that have (a) measured values of the relevant quantity, (b) \texttt{astroNN} uncertainty in $\logg<0.2$, which removes dwarf stars \citep{Leung19a}, and (c) relative \texttt{astroNN} distance uncertainty less than 20\,\%. For the $[\mathrm{Fe/H}]$ maps we further limit to stars with uncertainties $<0.05\,\mathrm{dex}$ and for the $[\mathrm{O/Fe}]$ maps we use the $[\mathrm{Fe/H}]$ cut as well as requiring the uncertainty in $[\mathrm{O/H}]$ to be $<0.05\,\mathrm{dex}$. 

\subsection{\emph{Gaia} data}

To determine the 3D kinematics of stars, we match the APOGEE data to the \gaia\ catalog \citep{GaiaDR2} using an angular matching radius of 2 arcseconds and record the proper motions, parallaxes, and their uncertainties (the parallaxes are only used to determine the systematic distance bias below).  We combine the proper motions with the line-of-sight velocities determined from the APOGEE spectra to obtain full 3D heliocentric velocities. We convert these to Galactocentric cylindrical coordinates using the same Galactic center and mid-plane distance as used for the positions above, and by using a solar motion of 11.1, 242., and $7.25\kms$ in the radial, rotational, and vertical direction, respectively \citep{Schoenrich10a,2012ApJ...759..131B}. 

\subsection{Distance biases}

\begin{figure*}
\begin{center}
\includegraphics[width=\textwidth]{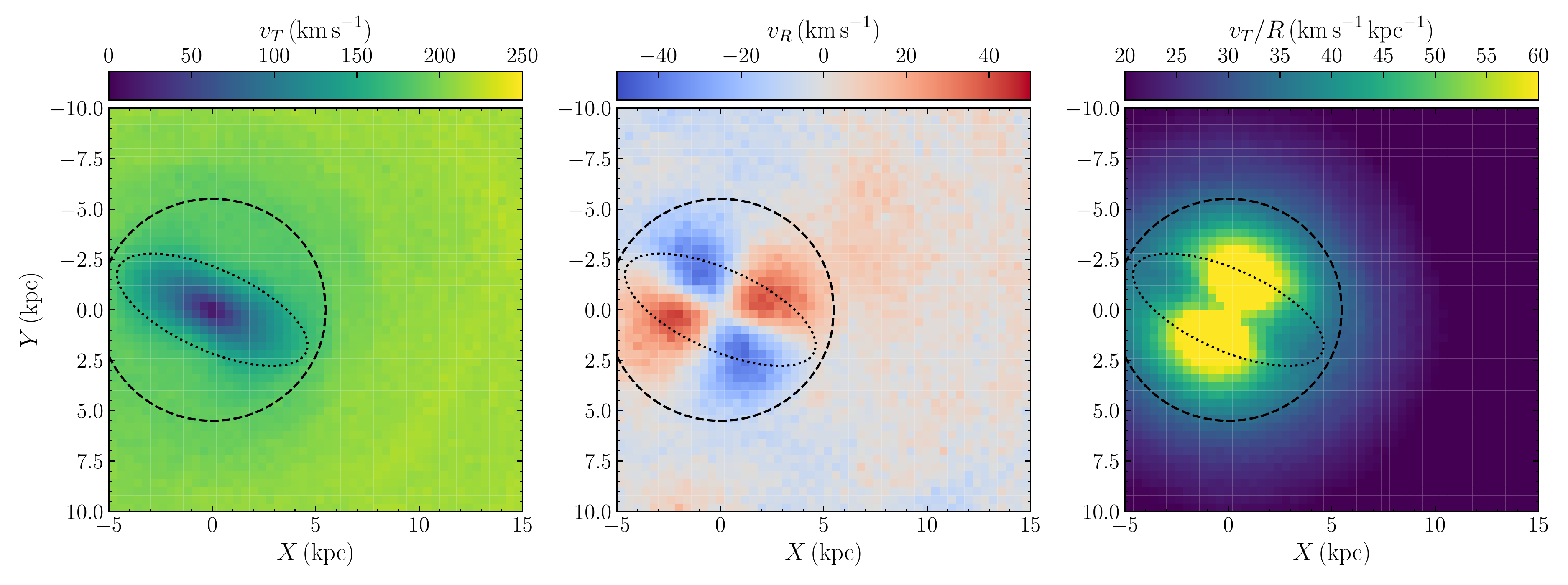}
\end{center}
\caption{Kinematics of stars in the bulge and disk of a barred, Milky-Way-like $N$-body simulation \citep{Kawata17a}, similar to Figure~\ref{fig:kinematics}. The bar's major axis in the simulation is oriented at $25^\circ$. The kinematic patterns in the simulation provide a remarkable match to those observed in Figure~\ref{fig:kinematics}.}\label{fig:barsim}
\end{figure*}

While the \texttt{astroNN} neural-network to determine distances returns precise distances to stars at large distances from the Sun, even a small relative bias will cause significant distortion of the observed structure at the large distance of the bar from the Sun. We determine whether there is a systematic bias in the distances determined using the neural network in the following manner. We select stars within 300 pc from the Galactic mid-plane based on their \texttt{astroNN} distances. To avoid outliers in the \gaia\ parallax measurements, we include stars with negative parallaxes only if they are larger than minus three times their uncertainty (but note that we do include negative parallaxes). We then assume that if \texttt{astroNN} determines that stars are at the same distance, that they truly are at the same distance, which is a reasonable assumption because a neural-network regression is very good at finding objects that have similar values of the output quantity (distance in this case) even if the output is biased. We then select stars at the same \texttt{astroNN} distance and determine the true distance $D$ of these stars by a maximum-likelihood fit to the parallax data, minimizing a robust objective function $\sum_i |\varpi_i-1/D|/\sigma_{\varpi,i}$, where $\varpi_i$ and $\sigma_{\varpi,i}$ are the parallax and its uncertainty of star $i$. We determine the uncertainty on $D$ using bootstrap resampling. In doing this fit, we add $52\,\mu\mathrm{as}$ to all \gaia\ parallaxes, the best-fit global parallax zero-point that we determined in a joint fit of the luminosity neural-network model and the zero-point \citep{Leung19b}.

The result of applying this procedure to stars in distance bins with a width of 200 pc is displayed in Figure~\ref{fig:astronn-dist-systematics}. We see that out to $4\kpc$, the \texttt{astroNN} distances have a small bias of 1\,\% in the sense of the \texttt{astroNN} distances being too large. But beyond $4\kpc$, the \texttt{astroNN} distances are too small by a fraction that grows to 10\,\% at $6\kpc$ and that stays constant at larger distances. The bias determined using stars towards the direction of the bulge alone agrees with the measurement using all stars. The source of this bias is likely related to the paucity of training data at the low-temperature, high-luminosity end that makes up the large-distance sample in the Galactic mid-plane: in regions of parameter space with few training data, regularized neural networks such as \texttt{astroNN} (which uses drop-out regularization) tend to regress to the mean, which in our case means that predicted luminosities are too low for these stars, thus underestimating their true distance.

We approximate the behavior shown in Figure~\ref{fig:astronn-dist-systematics} using the function
\begin{equation}
\begin{split}
    0.99  + 
    \frac{0.060}{\pi}\,&\left[\frac{\pi}{2}+\arctan\left(\frac{\tilde{D}/\mathrm{kpc}-4.35}{0.30}\right)\right]\\
     +
    \frac{0.055}{\pi}\,&\left[\frac{\pi}{2}+\arctan\left(\frac{\tilde{D}/\mathrm{kpc}-5.50}{0.15}\right)\right]\,,
\end{split}
\end{equation}
where $\tilde{D}$ is the raw \texttt{astroNN} distance. This function is shown as the smooth curve in Figure~\ref{fig:astronn-dist-systematics}. We correct all distances that we use in this paper using this function.

\section{The kinematics in the inner Galaxy and the bar's pattern speed}

In Figure~\ref{fig:kinematics}, we display the median cylindrical rotational and radial velocities of 73,189 stars observed by APOGEE and \gaia\ within 300 pc of the Galactic mid-plane as well as their rotational frequencies. These maps clearly show the signature of the Galactic bar: the rotational velocity map within 5 kpc from the center clearly demonstrates the expected bar-shape, with the major axis at $25^\circ$ clockwise from the Sun--Galactic center line, the radial velocity shows the tell-tale quadrupolar inwards and outwards movement on either side of the major axis, and the rotational frequency map displays the expected dip along the major axis. In Figure~\ref{fig:barsim}, we show these same maps for a barred, Milky-Way-like $N$-body simulation \citep{Kawata17a}, which illustrates the striking agreement between the observations and the expected bar kinematic pattern.

The kinematic maps from Figure~\ref{fig:kinematics} allow for a purely-kinematic, model-free determination of the bar's pattern speed, surface density profile, and shape through a novel application of the continuity equation. Our method is similar to the classical method of  \citet{Tremaine84a}, but because we employ full phase-space information, we do not have to assume that the pattern speed is constant over the bar region. Assuming that the surface density $\Sigma(X,Y,t)$ is constant in a frame rotating with the bar's pattern speed $\Omega_b$, $\Sigma(X,Y,t) \equiv \tilde{\Sigma}(R,\phi-\Omega_b t)$, the two-dimensional continuity equation becomes
\begin{equation}\label{eq:continuity}
    -\frac{\partial \Sigma}{\partial t} = \Omega_b\,\frac{\partial \tilde{\Sigma}}{\partial \phi} = \frac{1}{R}\,\frac{\partial (R\,v_R\,\tilde{\Sigma})}{\partial R} + \frac{1}{R}\,\frac{\partial (v_T\,\tilde{\Sigma})}{\partial \phi}\,,
\end{equation}
where $(R,\phi)$ are polar coordinates and all velocities are mean velocities in a small patch ($400\,\mathrm{pc}^2$ in Figure~\ref{fig:kinematics}). Near the major axis, $\partial \tilde{\Sigma}/\partial \phi \approx 0$, which using Equation~\eqref{eq:continuity} allows us to kinematically determine the radial surface-density profile along the major axis as
\begin{equation}\label{eq:continuity-major}
    \frac{\partial\ln\tilde{\Sigma}}{\partial R} = -\frac{1}{R\,v_R}\,\left(v_R + R\,\frac{\partial v_R}{\partial R} + \frac{\partial v_T}{\partial \phi}\right)\,.
\end{equation}
In practice, we assume that $\partial\ln\tilde{\Sigma}/\partial R$ is constant and find the best value by minimizing the median absolute difference between the left- and right-hand side of this equation. That we are able to determine the spatial profile of the inner Milky Way purely kinematically is important, because directly measuring the spatial profile by mapping the relative density of stars in our sample is complicated by the large and varying amount of interstellar extinction towards the inner Galaxy and the complex APOGEE selection function (e.g., \citealt{Bovy16b}).

\begin{figure}
\hspace{-0.5cm}\includegraphics[width=0.55\textwidth]{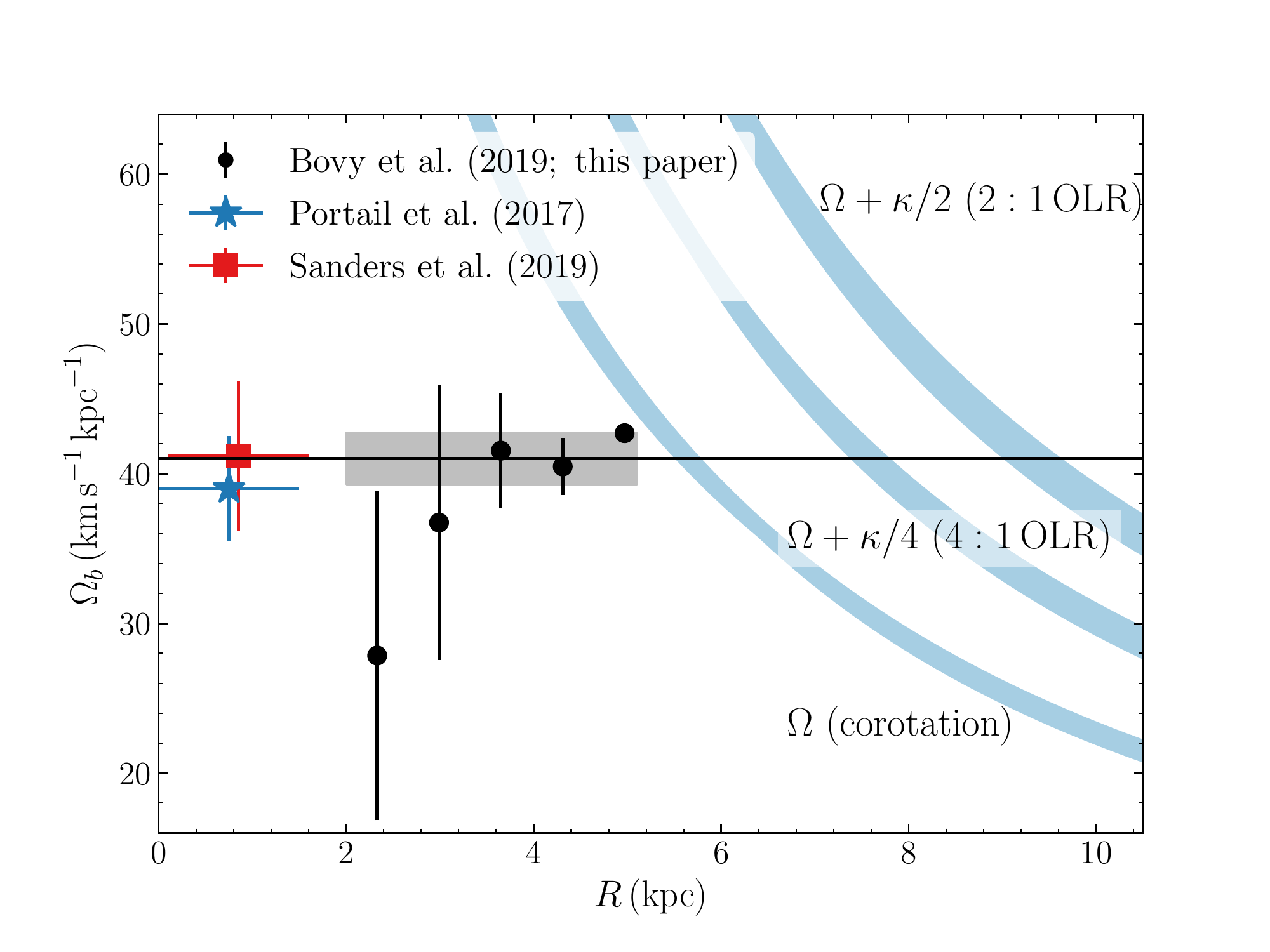}
\caption{Bar pattern speed measured through applying the continuity equation to the data in \figurename~\ref{fig:kinematics} as a function of radius (black points with errorbars). The horizontal line with the shaded uncertainty region displays our overall measurement of the bar's pattern speed of $\patspeed\kmskpc$. We also include the measurements of the bar's pattern speed using data at $R \approx1\kpc$ from \citet{Portail17a} and \citet{Sanders19a}. The curved, blue bands show the location of the bar's corotation and 4:1 and 2:1 outer Lindblad resonances in a range of Milky Way potential models. Our measurements combined with the inner-kpc determinations directly demonstrate that the entire inner Milky Way at $R < 5\kpc$ rotates as a solid body, i.e., as a stable bar.}\label{fig:barpattern}
\end{figure}

If we then assume that $\tilde{\Sigma}$ is stratified on similar, concentric ellipses with axis ratio $q$, all of the derivatives of $\tilde{\Sigma}$ at any $(R,\phi)$ in Equation~\eqref{eq:continuity} can be written in terms of $\partial\ln\tilde{\Sigma}/\partial R$ along the major axis and $q$ and we can determine $\Omega_b$ simply by combining the surface-density profile and a value of $q$ with the observed mean $v_R$ and $v_T$ velocities and their gradients. We re-write Equation~\eqref{eq:continuity} such that $\tilde{\Sigma}$ only appears as its logarithmic derivative along the major axis, $\partial\ln\tilde{\Sigma}/\partial \Ra$, which we kinematically determine using Equation~\eqref{eq:continuity-major}. We use $\Ra$ to denote the location along the major axis for a given point $(R,\phi)$. Assuming that the surface density is constant along concentric, similar ellipses with an axis ratio $q$ and a major axis along $\phi=0$, the function $\Ra(R,\phi) = R\,\sqrt{\cos^2\phi + \sin^2\phi/q^2}$. Therefore,
\begin{align}\label{eq:ellipse-bar-dSigmadR}
    \frac{\partial \ln \tilde{\Sigma}}{\partial R} & = \frac{\partial \ln \tilde{\Sigma}}{\partial \Ra}\,\frac{R}{\Ra}\,\left(\cos^2\phi+\frac{\sin^2\phi}{q^2}\right)\,,\\
    \frac{\partial \ln \tilde{\Sigma}}{\partial \phi} & = \frac{\partial \ln \tilde{\Sigma}}{\partial \Ra}\,\frac{R^2\,\sin2\phi}{2\Ra}\,\frac{1-q^2}{q^2}\,,\label{eq:ellipse-bar-dSigmadphi}
\end{align}
Working out the continuity equation in terms of these derivatives then gives the pattern-speed estimator
\begin{equation}
\begin{split}\label{eq:continuity-pattern}
    \Omega_b = \frac{v_T}{R}&+\frac{2q^2\,\Ra}{(1-q^2)\,R^2\,\sin2\phi\,\partial \ln \tilde{\Sigma}/\partial \Ra}\,\left(\frac{1}{R}\,\frac{\partial v_T}{\partial \phi}+\frac{v_R}{R}+\frac{\partial v_R}{\partial R}\right)
    \\ & +\frac{2\,v_R\,\left(q^2\,\cos^2\phi+\sin^2\phi\right)}{(1-q^2)\,R\,\sin2\phi}\,.
\end{split}
\end{equation}

The derivative of the continuity equation similarly allows us to determine the axis ratio $q$ itself kinematically. We take the $\phi$ derivative of the continuity equation with the substitutions from Equations~\eqref{eq:ellipse-bar-dSigmadR} and \eqref{eq:ellipse-bar-dSigmadphi}, eliminate $\Omega_b$ using the continuity equation, and set $\partial^2\ln\tilde{\Sigma}/\partial \Ra^2 = 0$, that is, we assume that the bar's surface density profile is exponential. This is a good assumption, because the Milky Way bar's outer density profile constrained using red-clump star counts is found to be exponential \citep{Wegg15a} and we see the same in the Milky-Way-like, barred $N$-body simulations below. This procedure leads to a complex equation that relates $q$ in a non-linear manner to $\partial\ln\tilde{\Sigma}/\partial \Ra$ and to the first and second derivatives of the observed kinematics $v_R$ and $v_T$. For each spatial bin, we can simply solve this equation by evaluating it on a grid of $q$ and finding the $q$ that best satisfies the equation. 

\begin{figure*}
\begin{center}
\includegraphics[width=\textwidth]{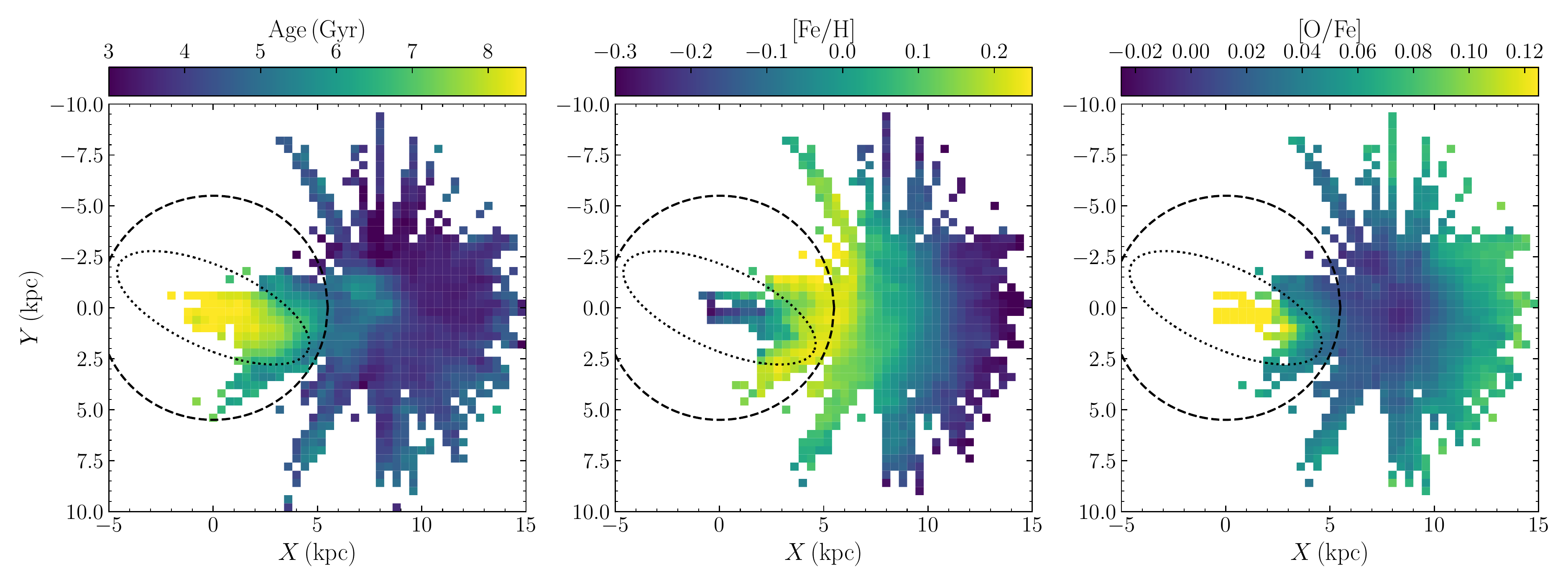}
\end{center}
\caption{Age, metallicity $ [\mathrm{Fe/H}]$, and oxygen-to-iron ratio  $[\mathrm{O/Fe}]$ of stars in the bulge and disk of the Milky Way. The circle and ellipse is as in Figure~\ref{fig:kinematics}. These maps for the first time reveal the Milky Way's bar in the distribution of ages and abundances, with a semi-major axis of $\approx5\kpc$ and a bar azimuthal angle of $\approx25^\circ$ that agrees with the kinematics in Figure~\ref{fig:kinematics} and with photometric studies.}\label{fig:ageabundances}
\end{figure*}

When applying Equation~\eqref{eq:continuity-major}, Equation~\eqref{eq:continuity-pattern}, and the equation for $q$ to the data and to the simulations below, we determine the first and second derivatives of $v_R$ and $v_T$ using simple finite-differencing of the maps in Figure~\ref{fig:kinematics}, which gives the gradients in rectangular coordinates, and convert these to the gradients in cylindrical coordinates. We then apply the estimators for $\partial\ln\tilde{\Sigma}/\partial \Ra$, $q$, and $\Omega_b$ to those pixels for which we can determine all of the relevant gradients within the region of interest (within 5$^\circ$ of the major axis when determining $\partial\ln\tilde{\Sigma}/\partial \Ra$ and beyond 5$^\circ$ of the major axis when determining $q$ and $\Omega_b$, at $2 < R/\mathrm{kpc} < 5$ in both instances). This gives an estimate for each pixel within the region of interest. We robustly combine these by taking the median and estimate the random uncertainty using the median absolute deviation, scaled by 1.4826 to convert the uncertainty to an effective Gaussian uncertainty.

Applying the procedure above to the data in Figure~\ref{fig:kinematics}, we determine that $\partial\ln\tilde{\Sigma}/\partial R = -1/(1.5\kpc)$ and $q=0.41\pm0.07$. Using Equation~\eqref{eq:continuity-pattern} then gives $\Omega_b = \patspeed\pm\patspeedranerr\kmskpc$. Applying the procedure to various barred, Milky-Way-like $N$-body simulations (see Appendix) recovers their pattern speeds to within this uncertainty and we can therefore be confident that the procedure is robust. Our method provides a measurement of the pattern speed at each spatial location and we can therefore investigate the spatial dependence of the measured pattern speed. Figure~\ref{fig:barpattern} displays the radial dependence of the pattern speed, determined by combining the measurements in $\Delta R = 0.33\kpc$ bins with at least 3 spatial pixels between $2\kpc$ and $5\kpc$; the combined measurement from the entire radial range is shown as the horizontal line with the gray-shaded error region. These radial measurements of $\Omega_b$, within their uncertainties, are consistent with the bar pattern speed being constant over $2\kpc < R < 5\kpc$.

In Figure~\ref{fig:barpattern}, we also include the location of the corotation resonance and of the 4:1 and 2:1 outer Lindblad resonances in the models of the Milky Way's gravitational potential from \citeauthor{Irrgang13a} (\citeyear{Irrgang13a}; model I), \citet{Bovy15a}, and \citet{McMillan17a} as implemented in \texttt{galpy} \citep{Bovy15a}. For a bar pattern speed of $\patspeed\kmskpc$, the 2:1 and 4:1 outer Lindblad resonance both strongly affect the kinematics of stars near the Sun \citep[e.g.,][]{Hunt18a,Hunt19a,Monari19a,Trick19a}.

The pattern speed that we determine depends systematically on the parameters of the transformation from heliocentric to Galactocentric coordinates and on the value of $\partial\ln\tilde{\Sigma}/\partial \Ra$ and $q$ that we determine. Varying the assumed value of the Sun's distance to the Galactic center between $8$ and $8.25\kpc$ leads to values of $\Omega_b$ in the range $38.5$ to $41\kmskpc$. Varying the Sun's rotational velocity with respect to the Galactic center---the most uncertain component and the one with the largest impact on our $\Omega_b$ determination---by $\pm5\kms$ changes $\Omega_b$ by $\pm1.25\kmskpc$. If rather than determining $\partial\ln\tilde{\Sigma}/\partial \Ra$ from the data, we set it to $-1/(1\kpc)$, we determine $\Omega_b = 42.5\kmskpc$; at the other extreme,  $\partial\ln\tilde{\Sigma}/\partial \Ra = -1/(2.5\kpc)$ leads to $\Omega_b = 36.5\kmskpc$. Similarly, setting $q$ to $0.3$ or $0.5$ gives $\Omega_b = 43$ and $38\kmskpc$, respectively. Considering these systematics, the recovery of the pattern speed in simulations to within $1.5\kmskpc$ (see Appendix) and the statistical uncertainty in our measurement, our data imply that $\Omega_b$ is contained in the interval $[36.5,42.5]\kmskpc$ with high confidence. To summarize this with a single value and uncertainty, we obtain $\Omega_b = \patspeed\pm\patspeederr\kmskpc$, but detailed comparisons with other determinations should take note of the systematics that lead to an asymmetric uncertainty interval .

\section{The formation and chemical evolution of the bulge}

\begin{figure*}
\begin{center}
\includegraphics[width=\textwidth]{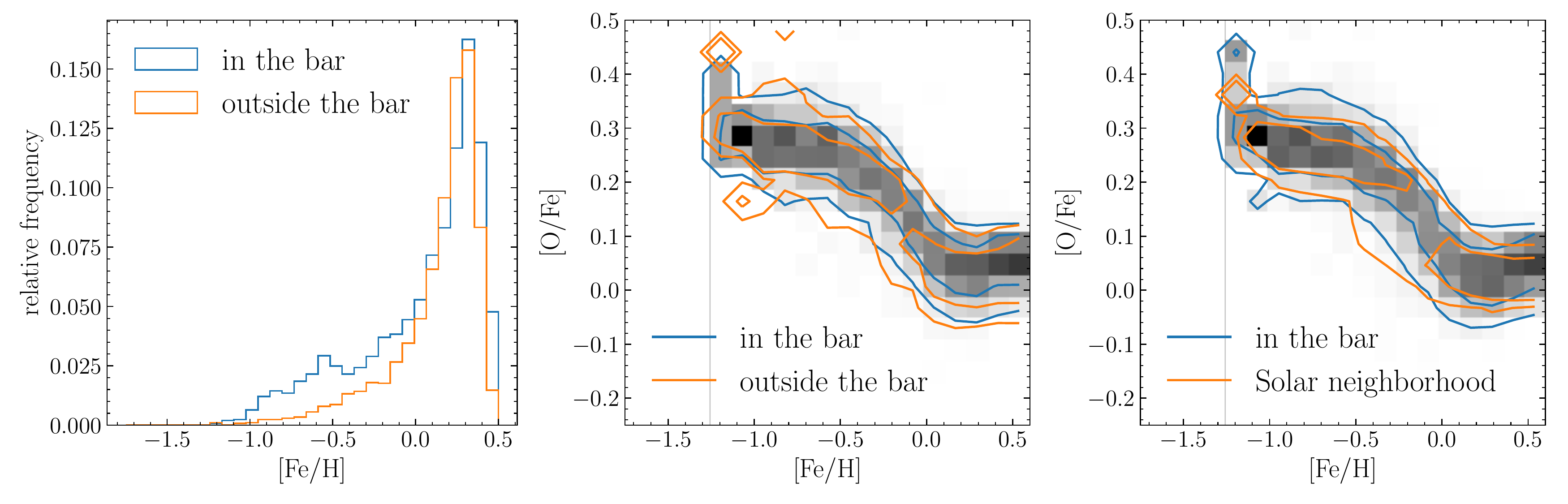}
\end{center}
\caption{Abundance distribution and chemical track of stars in the bar. The left panel displays the metallicity distribution, while the middle and right panels shows the abundance track in the $([\mathrm{Fe/H}],  [\mathrm{O/Fe}])$ plane. The grayscale in the middle and right panels shows the conditional distribution of $[\mathrm{O/Fe}])$ as a function of $ [\mathrm{Fe/H}]$ inside the bar, while the lines show 68\% and 95\% contours of this conditional distribution. We compare the abundance distribution of bar stars to that of inner-Galaxy stars outside the bar (left and middle panels) and to that of old stars near the Sun (right panel). The chemical evolution of all three populations shown is the same, but the bar consists of the older, metal-poor end of the evolutionary track.}\label{fig:abundancetracks}
\end{figure*}

In Figure~\ref{fig:ageabundances}, we present maps of the median age, iron abundance $[\mathrm{Fe/H}]$, and oxygen-to-iron ratio $[\mathrm{O/Fe}]$ of stars within 300 pc of the Milky Way's mid-plane; these maps are made with a total of 74,041, 67,308, and 62,970 stars, respectively. As in the kinematics, the bar clearly shows up in these maps, with a population that is old, metal-poor, and enhanced in oxygen compared to stars outside of the bar and in the disk. Contours of these maps clearly reveal a bar with a major axis inclined by $\approx25^\circ$ clockwise from the Sun--Galactic center line in agreement with the kinematic maps of Figure~\ref{fig:kinematics}. For example, fitting a model for the two-dimensional $[\mathrm{Fe/H}]$ that consists of concentric ellipses with a varying axis ratio, reveals a transition between an axis ratio of 0.45 near the center (in good agreement with the kinematically-determined axis ratio above) to an axis ratio of unity at $R \gtrsim 5\kpc$, therefore the size of the bar.

\begin{figure}
\begin{center}
\includegraphics[width=0.5\textwidth]{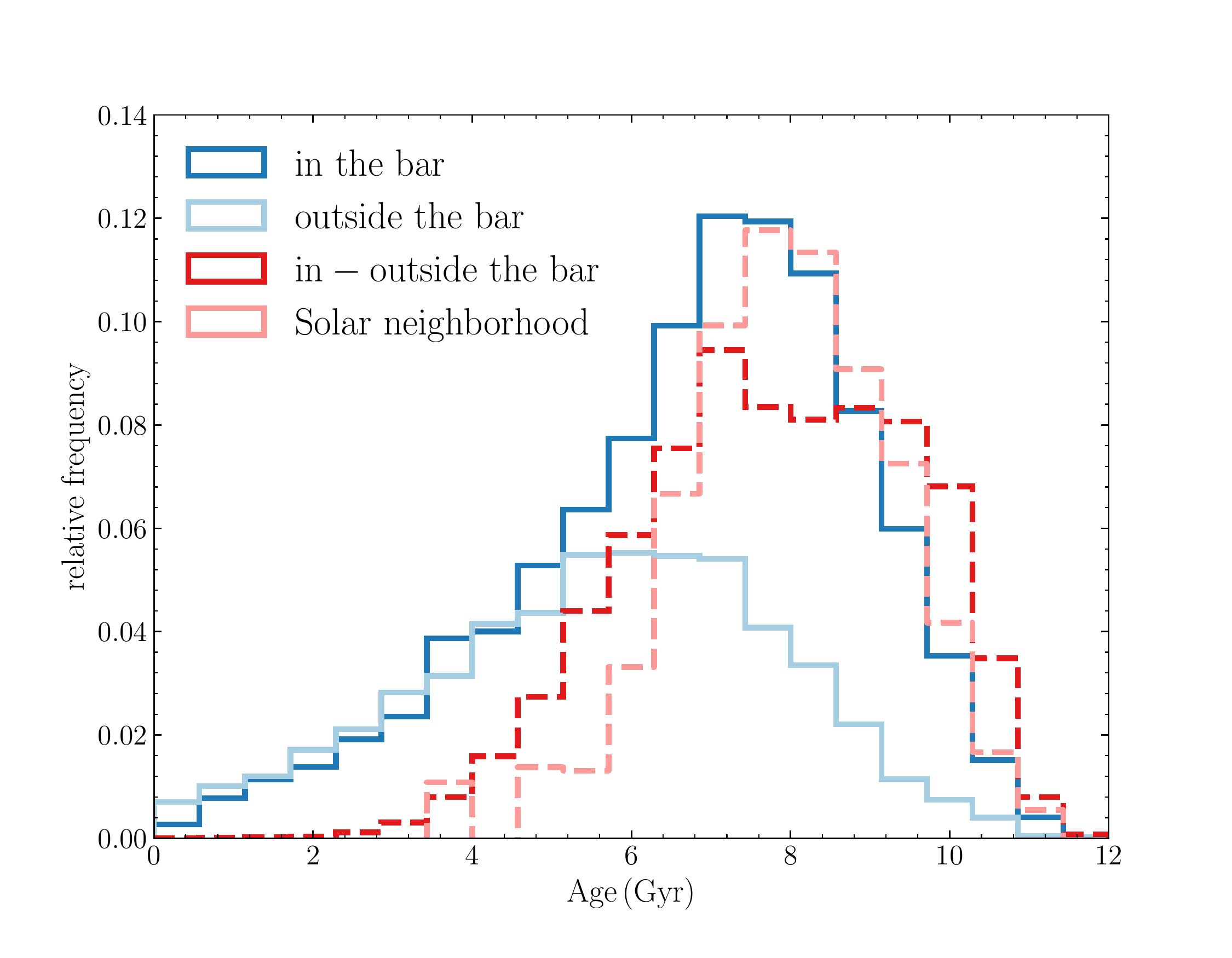}
\end{center}
\caption{Age distribution of giant stars in the bar compared to that of inner-Galaxy stars outside the bar and to that of similar old, Solar-neighborhood giants. The barred region consists of stars that are typically older than stars outside the bar, with the difference (purple, dashed curve) matching the age distribution of old disk stars (pink, dashed curve).  The bar therefore consists of stars formed during the early evolution of the Milky Way.}\label{fig:agedist}
\end{figure}

In Figure~\ref{fig:abundancetracks}, we display the metallicity distribution and $([\mathrm{Fe/H}],  [\mathrm{O/Fe}])$  track of stars in the inner Galaxy both inside and outside the bar. We select stars inside the bar as those stars within an ellipse with an axis ratio of $q=0.4$ and a semi-major axis of $5\kpc$ oriented at 25$^\circ$ clockwise from the Sun--Galactic center line (see Figure~\ref{fig:kinematics}); stars outside the bar are those with $R < 5\kpc$ that are outside of this ellipse. It is clear that stars in the inner Galaxy follow the same track, whether or not they are inside the bar, but the bar has an excess of more metal-poor stars with $[\mathrm{Fe/H}] < 0$ compared to stars outside the bar. Because the barred region contains a mix of bar members and members from the axisymmetric, inner-Galaxy component traced by the `outside the bar' distribution, this shows that the bar is primarily made up of metal-poor stars. In Figure~\ref{fig:agedist}, we similarly display the age distribution of stars inside and outside the bar in the central regions. The relation between the age distribution and the star-formation history is complex for the population of evolved, giant stars that we use as tracers. But, because the types of giants that we employ inside and outside the bar are the same, their relative formation epochs can be robustly determined. The bar is on average older than the rest of the inner Galaxy.

To connect the bar's evolution to that of the Galactic disk, we compare the bar's abundance and age distributions to those of the old stellar component in the Solar neighborhood, as traced by stars currently located more than 500 pc and less than 2 kpc from the mid-plane within $5 < R/\mathrm{kpc} < 10$, taking care to only select giants with temperatures below $4,600\,\mathrm{K}$ to match the temperatures of the stars observed in the bulge and avoid systematic biases in comparing ages and abundances determined for different stellar types. The right panel of Figure~\ref{fig:abundancetracks} compares the $([\mathrm{Fe/H}],  [\mathrm{O/Fe}])$  track of stars in the bar to that of old stars in the Solar neighborhood and we find them to be the same. In Figure~\ref{fig:agedist}, we compare the difference between the age distributions inside and outside the bulge to the age distribution of old Solar-neighborhood stars and find that they match remarkably well. Thus, the star-formation and chemical-evolution history of stars in the central barred region is the same as that of old stars near the Sun.

\section{Conclusion}

In this paper, we have presented the most complete maps of the kinematics, chemical abundances, and ages of stars in the Milky Way bulge and disk. For the first time, these maps directly reveal that the Milky Way disk--bulge system is barred not only in its density profile, but also in its kinematics, chemistry, and ages. These maps herald a new era in which we can unambiguously separate the properties of stars and their orbits in the bar from those in the inner disk, bulge, and stellar halo. 

We have used the maps of the kinematics of stars in the inner 5 kpc to determine the bar's pattern speed, radial profile, and shape in a purely kinematic manner using a simple, novel application of the continuity equation. In particular, we have determined that the bar's pattern speed $\Omega_b = \patspeed\pm\patspeederr\kmskpc$. Such a pattern speed places the bar's corotation radius at 5.1 to $5.9\kpc$ and the bar with size of $5\kpc$ therefore extends out to $\gtrsim90\,\%$ of its corotation radius, making the Milky Way's bar fast from a dynamical perspective. Prior measurements of the pattern speed, all based on incomplete data and/or involving complex and uncertain data modeling, range from $\approx30-40\kmskpc$ \citep{1999ApJ...524..112W,Rodriguez08a}  to $\approx60\kmskpc$ \citep{Englmaier99a}, with recent measurements \citep{Portail17a,Sanders19a} clustering around $40\kmskpc$, in good agreement with our simple determination here. Our measurement is the first determination of the bar's pattern speed in the bar's outer region, as previous measurements are largely based on data at $R \lesssim 2\kpc$, and we determine the radial dependence of the bar's pattern speed for the first time (see Figure~\ref{fig:barpattern}). The fact that our determination of the pattern speed at $2\kpc < R < 5\kpc$ is consistent with the recent measurements at $R < 2\kpc$ directly shows that the entire inner Milky Way rotates as a solid body, as expected for a rigid, stable bar.

The picture that emerges from the observations of the kinematics, chemistry, and ages of stars in the Milky Way presented in this work is that the bar formed as the early chemical evolution of the Milky Way's disk reached approximately solar abundances after a few Gyr of evolution. Studies of the age \citep{Mackereth17a}, abundance \citep{Bovy12a,Bovy16a}, and kinematics \citep{Bovy12c,Mackereth19a} of stars in the disk indicate that the early gas disk was thick due to turbulence, as also directly observed in more massive disk galaxies at redshift $\gtrsim 1$ \citep{ForsterSchreiber09a}, stabilizing it against bar formation. Direct measurements of the structure of different stellar disk components \citep{Bovy16a} show that as stars reached solar abundance ratios, the disk thickness had decreased to only a few hundred parsecs, at which time a bar instability could develop and the distribution of stars formed in the first few Gyr was re-arranged into a bar. Our direct observations of the chemical evolution of the bar show that this happened quickly. Chemical evolution in the inner disk then proceeded along the same chemical track to higher metallicities, while the bar rotated steadily as evidenced by our determination of its pattern speed, which is high for its size, indicating that it has not slowed down much. Our observations reveal the formation of the bar as a crucial epoch in the evolution of the Milky Way, separating the early turbulent epoch from the later, quiescent evolution \citep{Haywood18a}. A similar transition is hinted at \citep{Kraljic12a} in the cosmic evolution of the bar fraction \citep{Sheth08a} and disk turbulence \citep{Wisnioski15a}, but can now be directly observed in the Milky Way and placed into the wider disk-formation context.

\section*{Acknowledgements}

It is a pleasure to thank Francesca Fragkoudi, Ortwin Gerhard, David Nataf, Scott Tremaine, Benjamin Weiner, and the anonymous referee for helpful comments. J. Bovy and H. Leung received support from the Natural Sciences and Engineering Research Council of Canada (NSERC; funding reference number RGPIN-2015-05235) and from an Ontario Early Researcher Award (ER16-12-061). J. Hunt is supported by a Dunlap Fellowship at the Dunlap Institute for Astronomy \& Astrophysics, funded through an endowment established by the Dunlap family and the University of Toronto. J. Hunt performed part of this research at KITP with support from the Heising-Simons Foundation and the National Science Foundation (grant No. NSF PHY-1748958). We gratefully acknowledge the support of NVIDIA Corporation with the donation of a Titan Xp GPU used in this research. J.T. Mackereth acknowledges support from the ERC Consolidator Grant funding scheme (project ASTEROCHRONOMETRY, G.A. n. 772293).

Funding for the Sloan Digital Sky Survey IV has been provided by the Alfred P. Sloan Foundation, the U.S. Department of Energy Office of Science, and the Participating Institutions. SDSS-IV acknowledges
support and resources from the Center for High-Performance Computing at
the University of Utah. The SDSS web site is www.sdss.org.

SDSS-IV is managed by the Astrophysical Research Consortium for the 
Participating Institutions of the SDSS Collaboration including the 
Brazilian Participation Group, the Carnegie Institution for Science, 
Carnegie Mellon University, the Chilean Participation Group, the French Participation Group, Harvard-Smithsonian Center for Astrophysics, 
Instituto de Astrof\'isica de Canarias, The Johns Hopkins University, Kavli Institute for the Physics and Mathematics of the Universe (IPMU) / 
University of Tokyo, the Korean Participation Group, Lawrence Berkeley National Laboratory, 
Leibniz Institut f\"ur Astrophysik Potsdam (AIP),  
Max-Planck-Institut f\"ur Astronomie (MPIA Heidelberg), 
Max-Planck-Institut f\"ur Astrophysik (MPA Garching), 
Max-Planck-Institut f\"ur Extraterrestrische Physik (MPE), 
National Astronomical Observatories of China, New Mexico State University, 
New York University, University of Notre Dame, 
Observat\'ario Nacional / MCTI, The Ohio State University, 
Pennsylvania State University, Shanghai Astronomical Observatory, 
United Kingdom Participation Group,
Universidad Nacional Aut\'onoma de M\'exico, University of Arizona, 
University of Colorado Boulder, University of Oxford, University of Portsmouth, 
University of Utah, University of Virginia, University of Washington, University of Wisconsin, 
Vanderbilt University, and Yale University.

This work presents results from the European Space Agency (ESA) space mission Gaia. Gaia data are being processed by the Gaia Data Processing and Analysis Consortium (DPAC). Funding for the DPAC is provided by national institutions, in particular the institutions participating in the Gaia MultiLateral Agreement (MLA). The Gaia mission website is \url{https://www.cosmos.esa.int/gaia}. The Gaia archive website is \url{https://archives.esac.esa.int/gaia}.

\bibliographystyle{mnras}
\bibliography{refs}

\appendix

\section{Pattern speed determination for different simulations}

To test the determination of the pattern speed using the continuity equation, we apply it to three $N$-body simulations of barred disk galaxies that are set up to resemble the Milky Way. Figure~\ref{fig:barsim} displays the kinematics of stars within 300 pc of the mid-plane for the simulation of \citet{Kawata17a}, which clearly shows the same patterns as those observed in the data in Figure~\ref{fig:kinematics}. The other simulations are Target II and Target IV of \citet{Hunt13a}. The dark-matter halo is held fixed in all of these simulations and the bar's pattern speed is therefore quite stable, because the bar is unable to exchange angular momentum with the halo. We compute the orientation of the major axis of each simulation by fitting the surface-density profile of stars within 6 kpc from the center with a model where the density is constant along concentric, similar ellipses with an axis ratio that is a free parameter and a density profile along the major axis that is fit as a third-order polynomial. We then rotate all simulations such that their major axis is oriented at 25$^\circ$ clockwise from the Sun--Galactic center line to match the orientation of the Milky Way's bar. As we do for the data, we then first determine the surface-density profile $\partial\ln\tilde{\Sigma}/\partial \Ra$ along the major axis using Equation~\eqref{eq:continuity-major}, using the same cuts in angle and similar cuts in radius as for the data (we only use spatial bins along the major axis that are on the same side of the galaxy as the Sun; two of the simulations have slightly longer bars than the Milky Way's and for those we consider bins out to $R=6\kpc$). Comparing the kinematically-determined $\partial\ln\tilde{\Sigma}/\partial \Ra$ to the actual surface-density profile of the simulation shows excellent agreement. We then use this $\partial\ln\tilde{\Sigma}/\partial \Ra$ to determine $q$ with the equation for $q$ and then the pattern speed using Equation~\eqref{eq:continuity-pattern}, selecting spatial bins located more than $5^\circ$ away from the major axis, but additionally only using bins on the same side of the galaxy as the Sun and within $55^\circ$ of the major axis to approximately match the populated bins in the data. We determine the following pattern speeds: (a) $\Omega_b = 24.5\kmskpc$ for the \citet{Kawata17a} simulation with true $\Omega_B = 23.5\kmskpc$, (b) $\Omega_b = 32.2\kmskpc$ for the \citet{Hunt13a} Target II simulation with true $\Omega_B = 33\kmskpc$, and (c) $\Omega_b = 29.0\kmskpc$ for the \citet{Hunt13a} Target IV simulation with true $\Omega_B = 27.5\kmskpc$. The axis ratios that we determine kinematically agree to within a few percent with those determined near the end of the bar from a density fit to the simulated bars with varying axis ratio. It is clear that the assumption of a constant axis ratio does not limit our inference, because we are able to recover the simulations' pattern speed to high precision even though the simulations do not have constant axis ratios \citep{Hunt18a}.

\bsp	
\label{lastpage}
\end{document}